\documentclass[12pt,a4paper]{article}
\usepackage{amsmath,amssymb,url}
\usepackage{graphicx,tabularx}
\usepackage{array,dsfont}
\usepackage{color}

\newcommand\Rcite[1]{Ref.~\cite{#1}}

\makeatletter
\newif\if@preliminary
\@preliminaryfalse
\def\preliminary{\@preliminarytrue}
\def\preprintno#1{\def\@preprintno{#1}}
\def\address#1{\def\@address{#1}}
\def\email#1#2{\thanks{\tt #1@{}#2}}
\def\abstract#1{\def\@abstract{#1}}
\renewcommand\abstractname{ABSTRACT}
\newlength\preprintnoskip
\newlength\abstractwidth
\@titlepagetrue
\renewcommand\maketitle{\begin{titlepage}%
  \let\footnotesize\small
  \hfill\parbox{\preprintnoskip}{%
  \begin{flushright}\@preprintno\end{flushright}}\hspace*{1cm}
  \vskip 60\p@
  \begin{center}%
    {\Large\bf\boldmath \@title \par}\vskip 1cm%
    {\sc\@author \par}\vskip 3mm%
    {\@address \par}%
    \if@preliminary
      \vskip 2cm {\large\sf PRELIMINARY DRAFT \par \@date}%
    \fi
  \end{center}\par
  \@thanks
  \vfill
  \begin{center}%
    \parbox{\abstractwidth}{\centerline{\abstractname}%
    \vskip 3mm%
    \@abstract}
  \end{center}
  \end{titlepage}%
  \setcounter{footnote}{0}%
  \let\thanks\relax\let\maketitle\relax
  \gdef\@thanks{}\gdef\@author{}\gdef\@address{}%
  \gdef\@title{}\gdef\@abstract{}\gdef\@preprintno{}
}%
\def\@citex[#1]#2{\if@filesw\immediate\write\@auxout{\string\citation{#2}}\fi
  \def\@citea{}\@cite{\@for\@citeb:=#2\do
    {\@citea\def\@citea{,\penalty\@m}\@ifundefined
       {b@\@citeb}{{\bf ?}\@warning
       {Citation `\@citeb' on page \thepage \space undefined}}%
\hbox{\csname b@\@citeb\endcsname}}}{#1}}
\def\citerange{\@ifnextchar [{\@tempswatrue\@citexr}{\@tempswafalse\@citexr[]}}
\def\@citexr[#1]#2{\if@filesw\immediate\write\@auxout{\string\citation{#2}}\fi
  \def\@citea{}\@cite{\@for\@citeb:=#2\do
    {\@citea\def\@citea{--\penalty\@m}\@ifundefined
       {b@\@citeb}{{\bf ?}\@warning
       {Citation `\@citeb' on page \thepage \space undefined}}%
\hbox{\csname b@\@citeb\endcsname}}}{#1}}
\long\def\@makecaption#1#2{%
  \sbox\@tempboxa{#1: \emph{#2}}%
  \ifdim \wd\@tempboxa >\hsize
    #1: \emph{#2}\par
  \else
    \hbox to\hsize{\hfil\box\@tempboxa\hfil}%
  \fi
  \vskip\belowcaptionskip}
\def\fmslash{\@ifnextchar[{\fmsl@sh}{\fmsl@sh[0mu]}}
\def\fmsl@sh[#1]#2{%
  \mathchoice
    {\@fmsl@sh\displaystyle{#1}{#2}}%
    {\@fmsl@sh\textstyle{#1}{#2}}%
    {\@fmsl@sh\scriptstyle{#1}{#2}}%
    {\@fmsl@sh\scriptscriptstyle{#1}{#2}}}
\def\@fmsl@sh#1#2#3{\m@th\ooalign{$\hfil#1\mkern#2/\hfil$\crcr$#1#3$}}
\makeatother

\newcommand\whz{\textsc{Whizard}}

\newcommand\ltap{\
  \raise.3ex\hbox{$<$\kern-.75em\lower1ex\hbox{$\sim$}}\ }
\newcommand\gtap{\
  \raise.3ex\hbox{$>$\kern-.75em\lower1ex\hbox{$\sim$}}\ }

\newcommand\simge{\mathrel{%
   \rlap{\raise 0.511ex \hbox{$>$}}{\lower 0.511ex \hbox{$\sim$}}}}
\newcommand\simle{\mathrel{
   \rlap{\raise 0.511ex \hbox{$<$}}{\lower 0.511ex \hbox{$\sim$}}}}

\newcommand\be{\begin{equation}}
\newcommand\ee{\end{equation}}
\newcommand\bea{\begin{eqnarray}}
\newcommand\eea{\end{eqnarray}}
\newcommand\ba{\begin{array}}
\newcommand\ea{\end{array}}

\def\bq{\begin{equation}}
\def\eq{\end{equation}}
\def\ba{\begin{eqnarray}}
\def\ea{\end{eqnarray}}

\begin{document}

\date{\today}

\preprintno{DESY 16-032; MITP 16-021; SI-HEP-2016-07}

\title{Top Physics in WHIZARD}

\author{J\"urgen Reuter\email{juergen.reuter}{desy.de}$^{a,\ast}$, 
  Fabian Bach\email{fabian.bach}{t-online.de}$^b$,
  Bijan Chokouf\'{e} Nejad\email{bijan.chokoufe}{desy.de}$^a$,
  Andre Hoang\email{andre.hoang}{univie.ac.at}$^{c,d}$,
  Wolfgang Kilian\email{kilian}{physik.uni-siegen.de}$^e$,
  Maximilian Stahlhofen\email{mastahlh}{uni-mainz.de}$^f$,
  Thomas Teubner\email{thomas.teubner}{liverpool.ac.uk}$^g$,
  Christian Weiss\email{christian.weiss}{desy.de}$^{a,e}$
}

\address{\it%
$^a$DESY Theory Group, \\
  Notkestr. 85, D-22607 Hamburg, Germany
\\[.5\baselineskip]
$^b$European Commission, Eurostat, \\
2920 Luxembourg, Luxembourg
\\[.5\baselineskip]
$^c$
University of Vienna, Faculty of Physics, \\
Boltzmanngasse 5, 1090 Vienna, Austria
\\[.5\baselineskip]
$^d$
Erwin Schr\"odinger International Institute for Mathematical Physics,
University of Vienna, \\
Boltzmanngasse 9, 1090 Vienna, Austria
\\[.5\baselineskip]
$^e$
University of Siegen, Emmy-Noether Campus, \\
Walter-Flex-Str. 3, 57068
Siegen, Germany 
\\[.5\baselineskip]
$^f$
PRISMA Cluster of Excellence, Institute of Physics, Johannes Gutenberg
University, \\
Staudingerweg 7, 55128 Mainz, Germany 
\\[.5\baselineskip]
$^g$
University of Liverpool, Department of Mathematical Sciences, \\
Liverpool L69 3BX, U.K. 
\\[2\baselineskip]
$^\ast$ Talk presented at the International Workshop on Future Linear
Colliders (LCWS15), Whistler, Canada, 2-6 November 2015
}

\abstract{
In this talk we summarize the top physics setup in the event generator 
\textsc{Whizard} with a main focus on lepton colliders. This includes
full six-, eight- and ten-fermion processes, factorized processes and
spin correlations. For lepton colliders, QCD NLO processes for top
quark physics are available and will be discussed. A special focus is
on the top-quark pair threshold, where a special implementation
combines a non-relativistic effective field theory calculation
augmented by a next-to-leading threshold logarithm resummation with a
continuum relativistic fixed-order QCD NLO simulation.
}

\maketitle


\section{Top Physics in WHIZARD}

In this talk, the capabilities of the multi-purpose event generator
\whz{} focusing on top physics for lepton colliders are
covered. Though \whz{} is used by both LHC collaborations, ATLAS and
CMS, for top physics simulations, we concentrate here exclusively on
general issues and topics special for lepton colliders.

\whz{}~\cite{Kilian:2007gr} is a modular package, that contains it own
(tree-level) matrix element generator
\textsc{O'Mega}~\cite{Moretti:2001zz}. It uses recursive algorithms to
generate code avoiding all kinds of redundancies in the amplitudes from
the very beginning, for the Standard Model (SM) and in principle
arbitrary generalizations thereof. \whz{} contains a special module,
\textsc{CIRCE1/2}~\cite{Ohl:1996fi} that allows to simulate lepton
collider beam spectra including beam energy spectra, and also photon
collider options of linear lepton
colliders. \textsc{VAMP}~\cite{Ohl:1998jn} is \whz{}'s adaptive
multi-channel Monte Carlo integrator. The elaborate phase space
parametrization inside \whz{} is particularly suited for electroweak
productions at (but not only) lepton colliders. A massive support for
beyond the SM physics (supersymmetry, composite models, extra
dimensions, effective field theories) is
available~\cite{WHIZARD_BSM}. Beyond this, automatically generated BSM
models can be included via interfaces to external
tools~\cite{Christensen:2010wz}. \whz{} can handle (tree-level)
processes with six, eight or even ten fermions in the final state that
are needed to compute processes like $e^+e^- \to tt, tth$ completely off-shell.
Support for multi-threading with \texttt{OpenMP} and more elaborate techniques
to parallelize (SM) processes are provided~\cite{Nejad:2014sqa}.
\begin{figure}[htb]
	\centering
	\includegraphics[width=.45\textwidth]{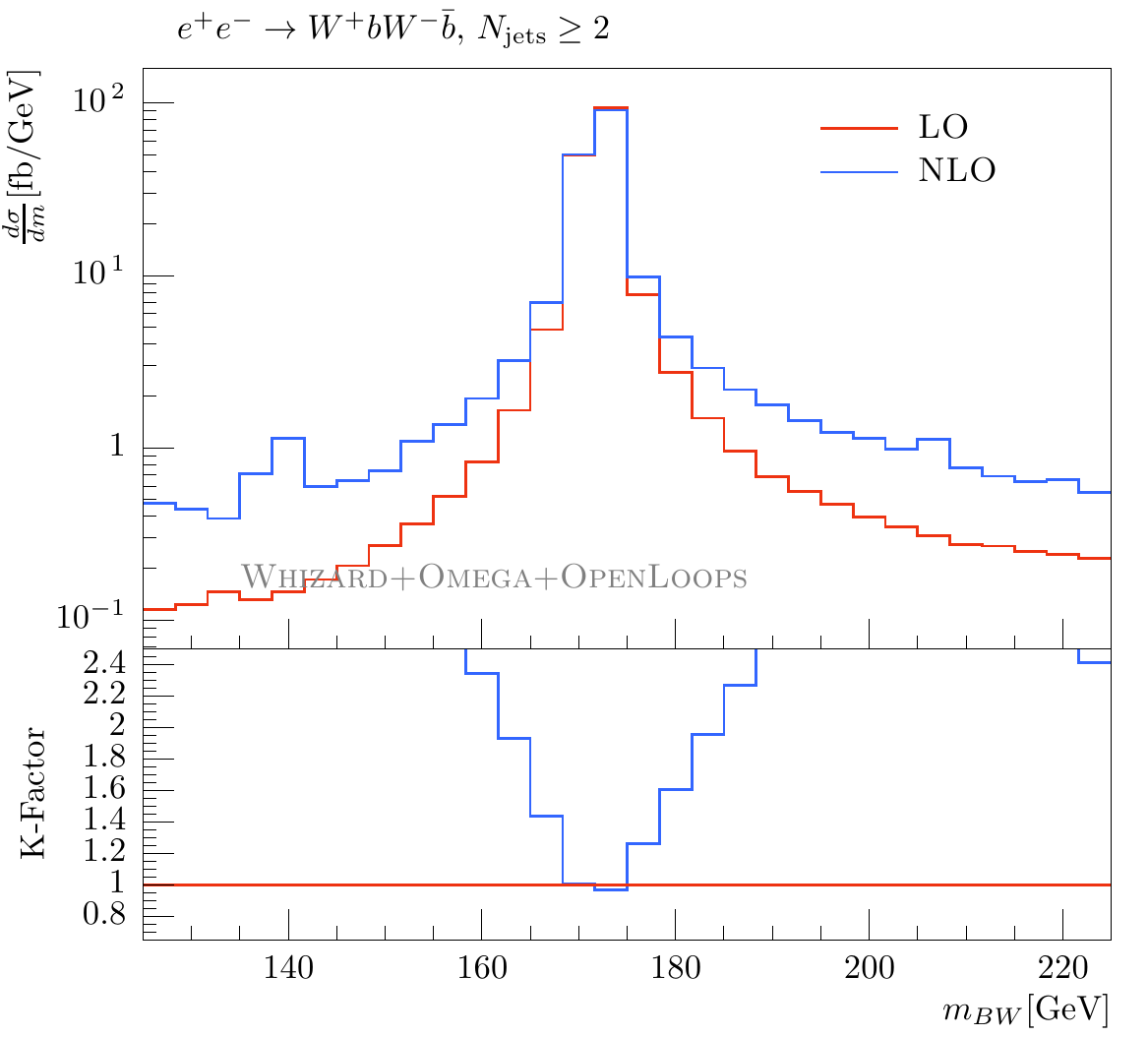}
	\includegraphics[width=.45\textwidth]{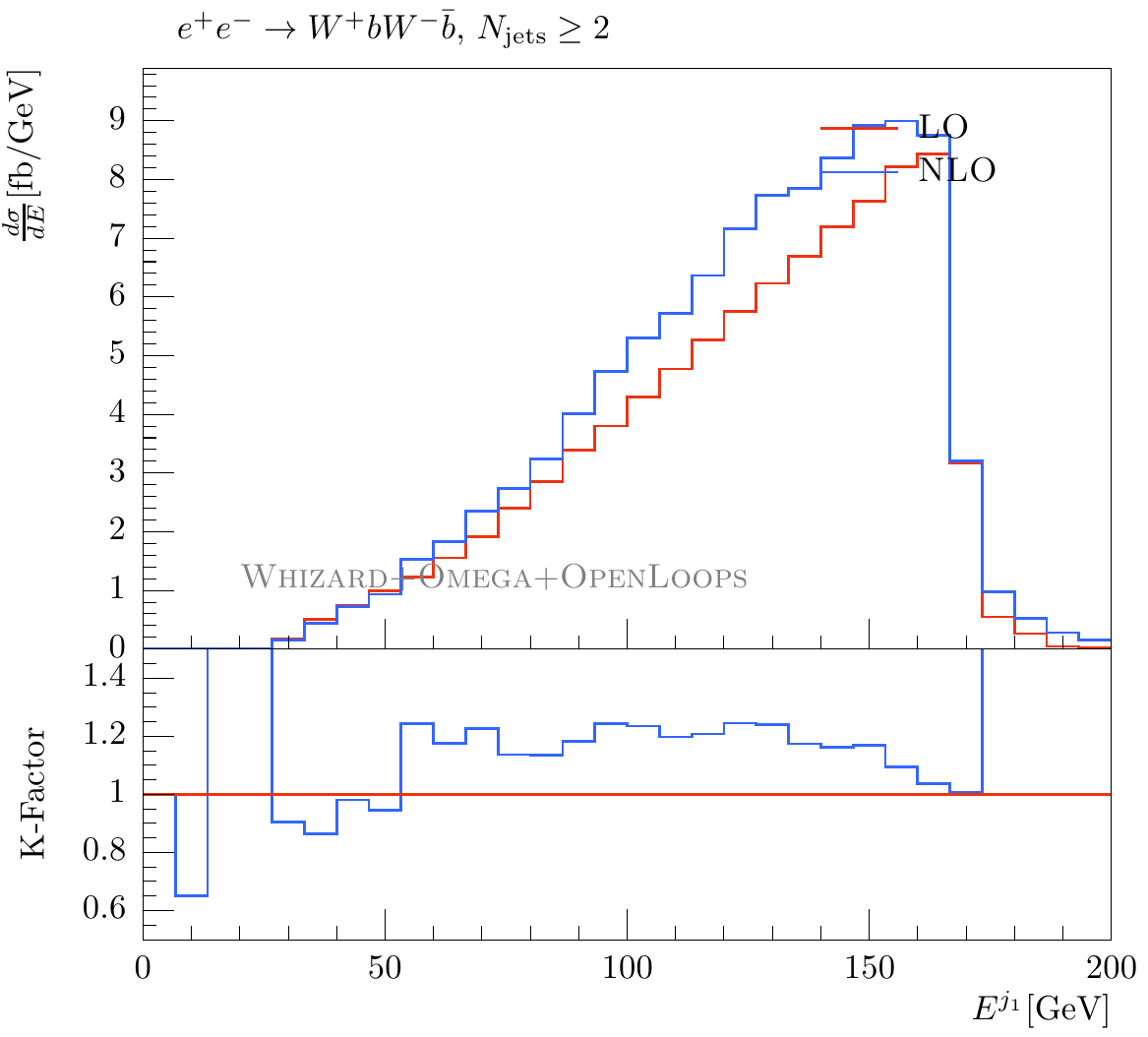}
  \caption{\label{fig:top-nlo}
    Resonant and non-resonant top pair production, $e^+e^- \to
    W^+ b W^- \bar b$ at a future ILC at $\sqrt{s} = 500$ GeV. The left
    plot shows the $Wb$ invariant mass distribution (where $b$ is the
    jet that contains the $b$ quark), while the plot on
    the right-hand side shows the energy of the hardest jet. Red lines
    are LO, blue ones NLO, respectively.}
\end{figure}
Besides full matrix elements, \whz{} allows to treat processes
factorized including full spin correlations, which for testing
purposes, however, can be switched off, or reduced to the classical
(diagonal) correlations. It is also possible to specify a particular
helicity of a decaying resonance. 

Top quarks are an important tool for searches for new
physics. Anomalous top quark couplings are implemented in \whz{} in an
effective-field theory setup including dimension-6 operators (this
also includes 4-fermion
operators)~\cite{Bach:2012fb,Bach:2014zca}. Furthermore,
flavor-violating top quark couplings have been implemented to allow
for the simulation of $t \to c$ transitions. 

To produce full events, \whz{} contains its own QCD parton shower
implementation, supporting both an analytical as well as a $p_T$
ordered shower~\cite{Kilian:2011ka}. To connect hard matrix elements
to the shower, \whz{} uses the color flow
formalism~\cite{Kilian:2012pz}. Hadronization, however, has to be done
using external tools. Precision state-of-art predictions
for SM processes should be at next-to-leading order (NLO) in the
strong coupling constant. After early attempts for the inclusion of
QED NLO corrections~\cite{Kilian:2006cj,Robens:2008sa} and QCD NLO
corrections~\cite{Binoth:2009rv,Greiner:2011mp}, \whz{} now allows for
the automatic generation of QCD NLO events using external virtual
one-loop matrix elements from either
\textsc{Gosam}~\cite{Cullen:2014yla} or OpenLoops~\cite{1111.5206}.
For a finite integration, soft- and collinear regions in an NLO
calculation have to be treated by a subtraction formalism. \whz{}
automatically generates FKS~\cite{Frixione:1995ms,Frixione:2009yq}
subtraction terms, i.e. the corresponding phase space mappings for the
soft- and collinear regions. As an example for top physics we show in
Figure~\ref{fig:top-nlo} differential distributions for the off-shell
top-quark production $e^+e^- \to W^+ W^- b\bar b$ at NLO QCD for
$\sqrt{s} = 500$ GeV. The left-hand side shows the $bW$ invariant mass
which is smeared out from the top peak mainly by the real radiation at
NLO, while the right-hand side shows the energy distribution of
hardest jet in the events. \whz{} also allows for a proper matching of
the fixed-order NLO calculation to the parton shower using the POWHEG
scheme, again for arbitrary (lepton-collider) processes.


\section{Top-antitop threshold}

The c.m. energy region around the top-antitop threshold, i.e. where
$\sqrt{s} \sim 2 m_t$, is of particular interest for experiments at a
future lepton collider like the ILC or CLIC. 
The measurement of the (total) cross section near threshold will allow
for a determination of the top mass in a theoretically well-defined
(short-distance) scheme and with unprecedented accuracy ($\delta m_t
\lesssim 100$ MeV) by fitting the theoretical prediction to the
resonance lineshape. 
Also other important (SM) parameters like the top decay width
(i.e. $V_{tb}$), the strong coupling $\alpha_s$ or the top Yukawa
coupling can be determined very precisely from the cross section close
to threshold~\cite{Seidel:2013sqa,Baer:2013cma}. 
Such measurements can in principle be fully inclusive, but in practice
due to experimental cuts and tagging in the final state, more
exclusive/differential observables might help to improve the precision
of the extracted top mass or increase the sensitivity to the parameter
of interest. 

Our aim is therefore to provide fully differential Monte Carlo
predictions for the top threshold region. 
Crucially, this requires the resummation of Coulomb singular terms
$\propto (\alpha_s/v)^n$ to all orders in perturbation theory, which
reflect the bound-state nature of the non-relativistic top-antitop
system. Here $v \sim \alpha_s \sim 0.1$ represents the relative
velocity of the top quarks close to threshold. 
In addition to the Coulomb singularities, large logarithms $\propto
\ln^n v$ can spoil the perturbative series and should be resummed. 
The resummation of all threshold enhanced terms can be systematically
carried out using a Schr\"odinger equation and the velocity
renormalization group within the non-relativistic effective field
theory vNRQCD~\cite{Luke:1999kz}. In vNRQCD the relevant dynamical
scales are the soft scale, given by the top momentum $\sim m_t v$, and
the ultrasoft scale, given by the kinetic energies of the tops $\sim
m_t v^2$. Particle modes with momenta of the order of the hard scale
$m_t$ (or bigger) have been integrated out. 

The decay of the top quarks -- predominantly into $W b$ -- plays a key
role for the prediction of top-antitop threshold production. 
The large decay width $\Gamma_t \gg \Lambda_{\rm QCD}$ effectively
serves as an infrared cut-off in the vNRQCD calculation and thus
allows for a perturbative description of the process. 
Upon resummation of the singular terms the normalized cross section
(R-ratio) close to threshold schematically takes the form 
\begin{align}
  R = \frac{\sigma_{t\bar{t}}}{\sigma_{\mu^+\mu^-}} \sim v \sum\limits_k \bigg(\!\frac{\alpha_s}{v}\!\bigg)^{\!\!k} \sum\limits_i (\alpha_s\,\ln\, v )^i \times \bigg\{\! 1 \, ({\rm LL});\;\alpha_s, v \, ({\rm NLL});\;\alpha_s^2,\, \alpha_s v,\, v^2\,({\rm NNLL});\ldots \! \bigg\},
\label{Rstruc}
\end{align}
where we have indicated the terms at leading-logarithmic (LL) order,
next-to-leading logarithmic (NLL) order, etc. 
The most up-to-date vNRQCD prediction of the total cross section has
reached NNLL~\cite{Hoang:2010gu,Hoang:2011gy,Hoang:2013uda}
precision. 

\subsection{Treatment inside the generator}

For the implementation in \whz{}, which was first discussed 
in~\cite{1411.7318}, we supplement the SM $tt\gamma$ and $ttZ$ (vector 
and axial-vector) vertices in the LO $e^+e^- \to W^+  W^- b\bar b$
Monte Carlo process with NLL (S- and P-wave) non-relativistic form
factors. As this is a modification of the plain standard model, a
special model, \texttt{SM\_tt\_threshold} is available for that
purpose in \whz{}.
The mentioned $tt\gamma$ and $ttZ$ form factors consist of a vertex
(Green) function and a Wilson coefficient, which is subject to
(velocity) RG running at NLL~\cite{Pineda:2001et,Hoang:2006ty}. 
The Green function is computed with the \textsc{Toppik}
code~\cite{Hoang:1999zc}, which numerically performs the Coulomb
resummation, and for the purpose of the threshold resummation is
shipped together with the \whz{} distribution.
For numerical stability and sufficient speed during integration and
event generation, we create interpolation grids for the form factor in
$\sqrt{s}$ and the square of the top three-momentum prior to the MC
integration. As we are aiming for differential cross sections, it is
important to also include the P-wave contributions, which only
represent a minor effect in the total cross section but are crucial to
describe e.g. the forward-backward asymmetry correctly. 
Besides providing fully differential predictions, the embedding in
\whz{} and \textsc{O'Mega} also has the advantage that the leading
effects from the interference with the non-resonant $W^+  W^- b\bar b$
background can be taken into account. 

A subtle but pivotal aspect of this computation is the top decay.
Per default, \textsc{O'Mega} attaches the tree-level decay of the top to the resummed production graph.
This implies that the LO width should be used in all top propagators
such that upon integration the total on-shell $t\,{\bar t}$ cross
section is approximately reproduced as $\Gamma_t \ll m_t$. 
On the other hand, to reconcile the prediction for the threshold with
the continuum in Section~\ref{ssec:matching}, we have to use the NLO width
everywhere. This in turn requires the NLO decay of the tops, also
in the non-relativistic computation. 
Furthermore, in combination with the LL resummed production, the
NLO correction to the decay can be an important NLL effect in
exclusive cross sections near threshold and should be taken into
account. It is particularly important when an additional hard gluon is 
resolved. We are currently working on a consistent implementation of
the NLO top decay using factorization. 
In this approach, we refrain from applying an on-shell projection, but
evaluate production and decay matrix elements with the mass set to the
top invariant mass. 
This allows to apply the factorization also \emph{below threshold} and
is similar to the procedure in \Rcite{1412.1828}, but without the boost
to equal invariant masses. 
An additional benefit of this implementation of the decay in the
threshold resummed process is that gauge invariance is guaranteed by
working with on-shell decays instead of a restricted set of Feynman
diagrams, i.e. signal diagrams. As in every resummed computation used
for event generation, this setup has full NLL+NLO accuracy only
for sufficiently inclusive observables. For arbitrarily exclusive
observables the precision is formally limited to LL+NLO, where the LL
is with respect to threshold resummation.

\begin{figure}[htb]
	\centering
	\includegraphics[width=30pc]{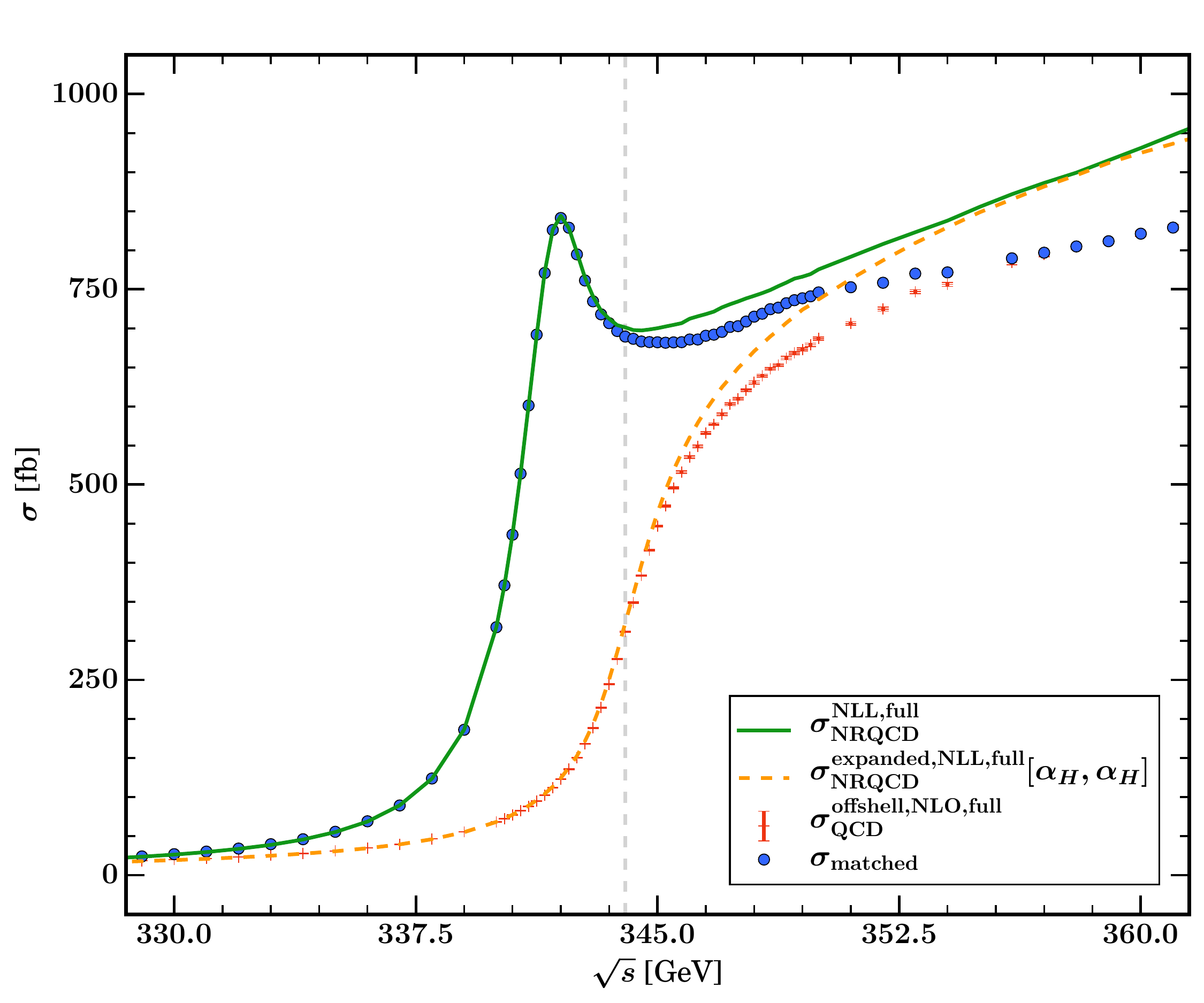}
  \caption{\label{fig:matching} Matching the NLL resummed threshold prediction to the fixed-order NLO QCD continuum for the total cross section of the process $e^+ e^- \rightarrow W^+bW^-b$.
    The blue dots show our (preliminary) result of the matching between the non-relativistic threshold and relativistic continuum region.
    The solid green line corresponds to the LO process with insertions of the
    NLL resummed form factors.
    The orange curve shows the same result with the form factors expanded
    to first order in $\alpha_s$ with relativistic (hard) scales, i.e.
    $\alpha_H\equiv\alpha(m_t)$.
    The red crosses represent the full relativistic fixed-order NLO result.}
\end{figure}

We have performed several cross checks of the numerical implementation
of the non-relativistic resummation in the \whz{} code. 
For example, we have verified that the analytic results, augmented
with relativistic corrections, are precisely reproduced for $\alpha_s
\to 0$ when using moderate cuts. 
We can also use analytic results as crosscheck in the threshold
region, i.e. where $v \to 0$ or $\sqrt{s} \to 2 m_t$: 
For the on-shell process, we have verified that \whz{} perfectly
reproduces the analytic result using on-shell form factors expanded to
$\mathcal{O}(\alpha_s)$ in the threshold region. 
Another important consistency check is the agreement of the prediction
using the expanded off-shell form factor with the full QCD NLO result
in the threshold region as shown in Figure~\ref{fig:matching}. 

\subsection{Matching to the relativistic continuum}
\label{ssec:matching}

Future lepton colliders might run at energies close to, but a bit off threshold (cf. e.g. the 380 GeV staging from the CLIC study group).
Therefore a smooth transition (matching), between the resummed threshold prediction and the fixed order prediction at large energies is required.
The necessary ingredients for this are:
the full $e^+ e^- \rightarrow W^+bW^-b$ NLO result, the threshold
resummed vNRQCD prediction, its expansion up to
$\mathcal{O}(\alpha_s)$ (with relativistic scale setting) and a
switch-off function that turns off the unphysical resummation
effects away from threshold.
For the NLO result, we can obtain the virtual amplitude conveniently via the BLHA interface from \textsc{OpenLoops}~\cite{1111.5206}.
The FKS subtraction in \whz{}~\cite{1510.02666} automatically
identifies the singular regions and adds and subtracts the necessary
terms in the real and virtual components. 
We then add the non-relativistic resummed process and, to avoid double
counting, subtract the non-relativistic NLO terms with the hard scale,
cf. the orange curve in Figure~\ref{fig:matching}. 
As noted already above, {this expanded form factor} gives a very good
approximation of the full process as it contains the dominant terms
close to threshold. 

Finally, we have to cure the problem that the resummed prediction
keeps growing arbitrarily with $\sqrt{s}$, which indicates the
break-down of the non-relativistic approximation and is seen by the 
rise of the green curve in Figure~\ref{fig:matching}.  
This is done by multiplying the relevant couplings with a switch-off
function that smoothly approaches zero  as one moves away from
threshold. One has some freedom in the definition of this function as
well as the decision where to switch off the resummation, i.e. at
which energies one does not trust the resummed results anymore. 
This freedom has to be treated as an uncertainty that should
eventually be combined with the scale uncertainties (and other
potential error sources) for a reliable theory error estimate in the
intermediate region. Given the rather smooth transition between the
threshold and continuum region in the matched prediction of
Figure~\ref{fig:matching}, we however expect this error to be relatively
small for the total cross section. In summary, the procedure described
above allows us to consistently add the terms beyond NLO from the NLL
threshold resummation to the full NLO QCD result in \whz{}.


\section{Summary}

In this talk, we presented the current status of top quark physics
inside the event generator \whz{} with a special emphasis on lepton
colliders. Two main ongoing projects are the general automation of
(QCD) fixed-order NLO corrections for SM processes and a proper
matching of the continuum off-shell top pair production with the
non-relativistic resummed corrections at the top threshold.

\section*{Acknowledgments}

JRR wants to thank the organizers for a great conference in the
Canadian wilderness.


\end{document}